\documentclass[conference]{IEEEtran}
\IEEEoverridecommandlockouts
\usepackage{cite}
\usepackage{orcidlink}
\usepackage{subcaption} % Add this to the preamble
\usepackage{float}
\usepackage{amsmath,amssymb,amsfonts}
\usepackage{algorithmic}
\usepackage{mathtools,amssymb}

\usepackage{cuted}
\hypersetup{%
    pdfborder = {0 0 0}
}
\usepackage{graphicx}
\usepackage{textcomp}
\usepackage{xcolor}
\def\BibTeX{{\rm B\kern-.05em{\sc i\kern-.025em b}\kern-.08em
    T\kern-.1667em\lower.7ex\hbox{E}\kern-.125emX}}
\begin{document}

\title{X-Band UAV-enabled Integrated Sensing and Communications for Vehicular Networks}
\author{\IEEEauthorblockN{Remon Polus \orcidlink{0000-0002-5527-0265}}
\IEEEauthorblockA{\textit{Department of Computer and Software Engineering} \\
\textit{Polytechnique Montréal}\\
Montréal, Canada \\
remon.polus@polymtl.ca }
\and
\IEEEauthorblockN{Soumaya Cherkaoui \orcidlink{0000-0001-6140-770X}}
\IEEEauthorblockA{\textit{Department of Computer and Software Engineering} \\
\textit{Polytechnique Montréal}\\
Montréal, Canada \\
soumaya.cherkaoui@polymtl.ca }
}

\maketitle
\begin{abstract}
Uncrewed aerial vehicles (UAVs) are increasingly considered as aerial platforms capable of providing both sensing and communication services, representing a promising paradigm for intelligent transportation systems.
This paper investigates the optimal time allocation for a UAV-enabled integrated sensing and communication (ISaC) system operating in the X-band for vehicular networks.
We analyze the trade-off between sensing accuracy and communication performance under practical UAV constraints and fading effects, considering both single-shadowing and double-shadowing channel models.
An optimization framework is developed to allocate time between sensing and communication while guaranteeing minimum communication rates and sufficient sensing reliability.
Simulation results demonstrate adaptive time allocation strategies, highlighting how UAV-to-ground channel conditions and target distances influence the balance between sensing and communication in smart mobility scenarios.

\end{abstract}
\begin{IEEEkeywords} 6G, Integrated sensing and communications (ISaC), uncrewed aerial vehicles (UAVs), time allocation.
\end{IEEEkeywords}
\section{Introduction}

\IEEEPARstart{T}{he} development of sixth-generation (6G) networks is progressing through international collaboration, with the IMT-2030 vision providing critical recommendations that extend the capabilities of IMT-2020 \cite{liu2025itu}.
Among the innovations highlighted in this vision, integrated sensing and communication (ISAC) has emerged as a fundamental technology for 6G.
ISAC facilitates the simultaneous provision of high-throughput communication and high-precision sensing, making it essential for a wide array of next-generation applications \cite{meng2024cooperative}.
Initially proposed to enable the shared use of frequency bands traditionally reserved for radar systems \cite{zheng2019radar}, ISAC offers two key advantages: integration gain, achieved through the joint use of resources such as spectrum, energy, and hardware, and coordination gain, which enhances overall system performance by fostering mutual support between sensing and communication functions \cite{dong2022sensing}.
These features position ISAC as a key technology that facilitates rapid in-situ sensing, which can be leveraged to improve mobility management and optimize handover decisions in high-mobility scenarios for vehicular networks \cite{aman2025isac}.

Uncrewed aerial vehicles (UAVs) further enhance the potential of ISAC by providing flexible and dynamic coverage for both communication and sensing.
Their mobility, coupled with the ability to operate at varying altitudes, enables real-time data collection and high-resolution traffic monitoring, incident detection, and data collection for intelligent transportation systems and smart cities \cite{mu2023uav}.
Leveraging their elevated altitude and controllable 3D mobility, UAVs-enabled ISAC can achieve wider sensing coverage, more flexible deployment, and better communication reliability compared to terrestrial systems \cite{pang2024dynamic,liu2020opportunistic}.
%These wide telecommunications applications of UAVs have prompted researchers to develop advanced fading channel models that accurately describe the fluctuations in communication links between UAVs and ground nodes \cite{bithas2020uav}, \cite{bithas2019new}.
%These models, based on Nakagami-$m$ and inverse-gamma distributions, effectively capture multipath and shadowing effects.
%The models were further validated through empirical measurements.

Recent studies on UAV-enabled ISAC have advanced resource allocation strategies to optimize system performance.
\cite{liu2024secure} focused on maximizing the secrecy rate by jointly optimizing user scheduling, transmit power allocation, and UAV trajectory, while \cite{zhang2024secure} explored secure transmissions in intelligent reflecting surface (IRS)-aided UAV-enabled ISAC. 
Other studies analyzed UAV trajectory design, optimizing both maneuvering and beamforming \cite{jing2024isac}, and \cite{meng2022throughput} worked on throughput maximization by jointly optimizing beamforming, user association, sensing time allocation, and UAV trajectories.
Energy efficiency is a critical challenge in UAV operations.
\cite{khalili2023energy} tackled this by optimizing UAV trajectory, velocity, and time allocation to minimize power consumption. 
Similarly, \cite{zhu2024resource} investigated energy-efficient resource allocation in UAV swarms, ensuring sustainable operations.

While most research has focused on transmit power allocation to improve system efficiency, less attention has been given to time allocation between sensing and communication signals, an area that remains underexplored.
Optimizing time allocation in an ISAC system is crucial, as the duration directly influences both sensing and communication efficiency.
Additionally, many studies assume a line-of-sight channel between UAVs and ground nodes, which may not always be realistic due to environmental conditions, suggesting the need for more robust channel models in future research.

This paper proposes a UAV-enabled ISAC system optimizing time allocation between communication and sensing modes.
The 10.05–10.5 GHz band, part of the X-band (8–12 GHz), falls within the Super High Frequency (SHF) range and is increasingly considered for 5G-Advanced deployment as a midband communication band \cite{testolina2024sharing}.
Known for its low atmospheric attenuation, this frequency range is ideal for airborne radar systems \cite{moses2014uav}. 
Its shorter wavelengths enable compact, high-performance antennas, making it well-suited for space-constrained airborne applications, highlighting its potential for UAV-enabled ISAC systems.
More specifically, the contributions of the paper are as follows:
\begin{itemize}
\item A novel ISAC system for airborne vehicles is proposed, which incorporates both small-scale fading and shadowing effects on the links between the UAV and the target.
\item An optimal time allocation problem is formulated, with constraints designed to ensure both target sensing and communication performance are met. 
\item Simulation results show that the proposed optimal approach not only satisfies the required communication and sensing performance but also achieves greater power efficiency compared to the equal time allocation method. 
\end{itemize}
The rest of this paper is organized as follows:
Section \ref{S2} introduces the system model.
The time allocation problem is formulated, and the solution approach is detailed in Section \ref{S3}.
Section \ref{S4} provides a thorough discussion of the simulation results.
Finally, Section \ref{S5} offers the conclusion.
    
\section{System Model}
\label{S2}
\begin{figure}[!t]
\centering
\includegraphics[width=\columnwidth]{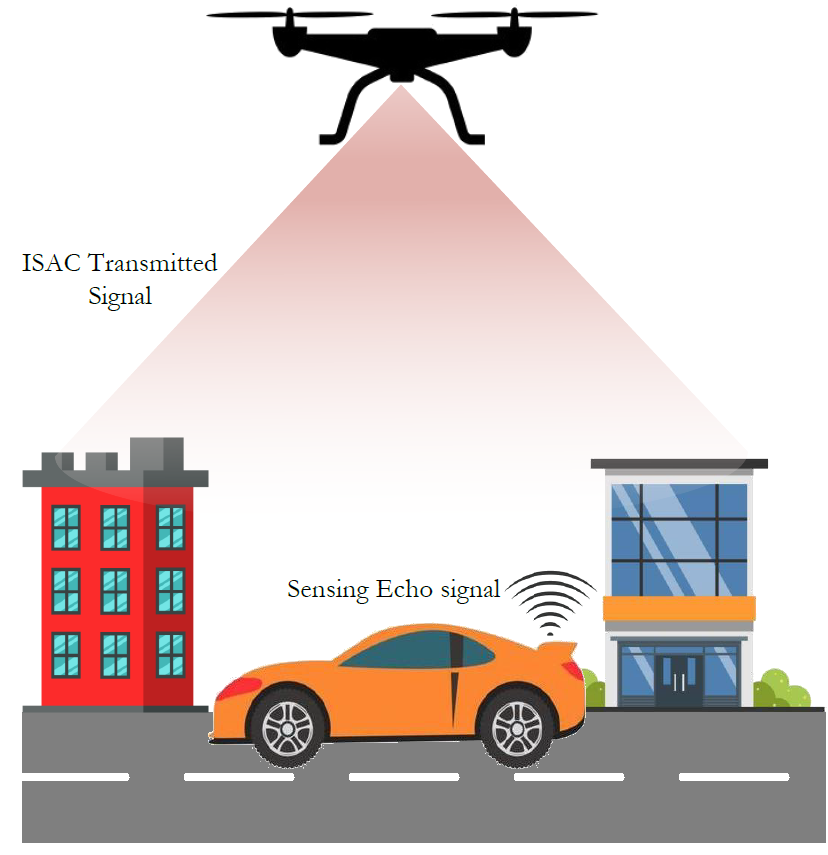}
\caption{UAV-enabled ISaC system model.}
\label{Figure 01}
\end{figure}

As shown in Fig. \ref{Figure 01}, a UAV transmits an ISaC signal to a moving vehicle at distance $d$ and velocity $v$. The UAV detects the vehicle by analyzing the echo signal received at its receiver, while the vehicle’s user equipment receives the data communication. 
The total ISaC signal duration, $T$, is split into two subframes: sensing mode, $T_s = \theta T$, and communication mode, $T_c = (1-\theta) T$, where $0 \leq \theta \leq 1$.

\subsection{Communications Mode}  

The received signal-to-noise ratio (SNR) at the user equipment is given by  
\begin{equation}
\label{eq01}
    \gamma_c=\frac{P_c|h|^2d^{-\eta}}{N_oB_c},
\end{equation}  
where \( B_c \) denotes the allocated bandwidth for communication, \( N_o \) represents the noise power spectral density, \( h \) is the fading coefficient between the UAV and the ground vehicle, \( \eta \) is the path loss exponent, and \( P_c \) is the transmission power.  

Based on Shannon’s capacity theorem \cite{shannon1948mathematical}, the maximum achievable communication rate can be expressed as  
\begin{equation}
\label{eq02}
    \mathbb{C}_c=\frac{B_c}{\ln2}\ln\left(1+\gamma_c\right).
\end{equation}  
The fading term \( |h|^2 \) follows a probabilistic model based on either the double-shadowing (DS) or single-shadowing (SS) conditions, as discussed in the following subsections.  

\subsubsection{\textbf{Double-Shadowing (DS) Communication Scenario}}  

In this case, the UAV serves as the transmitter, communicating with a receiver located on the ground. The channel undergoes independent shadowing effects at both ends of the link due to the spatial separation of the UAV and the ground node. The received SNR \( \gamma \) can be modeled as  \cite{bithas2020uav}.
\begin{equation}
\label{E01}
\gamma = N_1^2 I_1 N_2^2 I_2,
\end{equation}  
where:  

\begin{itemize}
    \item \( N_j \) (\( j \in \{1,2\} \)) denotes the multipath fading coefficient, following a Nakagami-\( m \) distribution \cite{nakagami1960m}, with the probability density function (PDF)  
    \begin{align}
    \label{E02}
    f_{N_j^2}\left(x\right)=\frac{m_j^{m_j} x^{m_j-1}}{\Omega_j^{m_j} \Gamma\left(m_j\right)}e^{-\frac{m_jx}{\Omega_j}}, 
    \quad x>0,
    \end{align}  
    where \( m_j \) represents the fading severity, \( \Omega_j \) is the scale parameter, and \( \Gamma(\cdot) \) is the Gamma function.  

    \item \( I_j \) captures the shadowing effect and follows an inverse-gamma (IG) distribution \cite{witkovsky2001computing}, with its PDF given by  
    \begin{align}
    \label{E03}
    f_{I_j}\left(x\right)=\frac{\overline{\gamma}_j^{\alpha_j}}{x^{\alpha_j+1} \Gamma\left(\alpha_j\right)}e^{-\frac{\overline{\gamma}_j}{x}}, 
    \quad x>0,
    \end{align}  
    where \( \alpha_j > 1 \) indicates the shadowing intensity, and \( \overline{\gamma}_j \) is the scale parameter.  
\end{itemize}

The resulting PDF of the received SNR is given by \cite{bithas2020uav}.
\begin{equation}
\label{E04}
\begin{split}   
f_{\gamma}\left(\gamma\right) &= 
\frac{\mathbb{S}_{DS}}{\gamma}
G_{2,2}^{2,2}\left(\frac{m_1m_2}{\overline{\gamma}}\gamma
\bigg|
\begin{matrix}  1-\alpha_2,1-\alpha_1  \\ m_1,m_2  \end{matrix}\right),
\end{split}
\end{equation}  
where  
\[
\mathbb{S}_{DS} = \frac{1}{\Gamma\left(m_1\right)\Gamma\left(m_2\right)\Gamma\left(\alpha_1\right)\Gamma\left(\alpha_2\right)}
\]
and \( G_{p,q}^{m,n}(\cdot|\cdot) \) is the Meijer \( G \)-function, which can be computed using built-in MATLAB functions [\citenum{gradshteyn2014table}, eq. (9.301)].
\subsubsection{\textbf{Single-Shadowing (SS) Communication Scenario}}
In the event of a single shadowing region located near one of the Tx or the Rx, the channel is categorized as a single-shadowing (SS) channel whose received SNR can be modeled as \cite{bithas2020uav}.
\begin{equation}
\label{E07}
\gamma=N_1^2 N_2^2 I.
\end{equation}
As shown in \cite{bithas2020uav}, the PDF of the received SNR can be formulated as.
\begin{equation}
\label{E08}
\begin{split}
f_{\gamma}(\gamma)&=
    \frac{\mathbb{S}_{SS}}{\gamma}
    G_{1,2}^{2,1}\left(\frac{m_1m_2}{\overline{\gamma}}\gamma\bigg|
    \begin{matrix} 1-\alpha  \\ m_1,m_2  \end{matrix}\right),
\end{split} 
\end{equation}
where $\mathbb{S}_{SS} = \frac{1}{\Gamma\left(m_1\right)\Gamma\left(m_2\right)\Gamma\left(\alpha\right)}$.

\subsection{Sensing Mode}

The UAV is equipped with a continuous wave frequency modulated (CWFM) radar system for precise range and velocity estimation of ground vehicles, classified as an airborne ground surveillance radar.
As outlined in \cite{gaur2024optimal}, the sensing capacity is given by:
\begin{equation} 
\label{eq2.07}
\begin{split} 
\mathbb{C}_s &= \frac{1}{\ln2} \ln \left( 1 + \frac{8B_sT_sR_mV_m}{c\lambda} \right), 
\\ &= \frac{1}{\ln2} \ln \left( 1 + \gamma_s \theta \right), 
\end{split} 
\end{equation}
where $\gamma_s = \frac{8B_sR_mV_mT}{c\lambda}$, with $B_s$ being the sensing bandwidth, $R_m$ the maximum detectable range, $V_m$ the maximum velocity, $c$ the speed of light, and $\lambda$ the wavelength.

To ensure optimal radar performance, particularly in minimizing false alarms and maximizing detection probability, the echo signal-to-noise ratio (SNR) is critical.
For a CWFM radar \cite{instruments2017programming}, the minimum required echo SNR to detect $R_m$ is:

\begin{equation}
\label{eq2.08}
\delta_{min} = \frac{P_sG_tG_r\lambda^2\sigma T_s}{(4\pi)^3R_m^4k_B T_o},
\end{equation}
where $P_s$ is the sensing transmit power, $G_t$ and $G_r$ are the transmitter and receiver antenna gains, $\sigma$ is the radar cross section, $k_B$ is Boltzmann's constant, and $T_o$ is ambient temperature.
Rearranging (\ref{eq2.08}), the sensing fraction $\theta$ must satisfy:
\begin{equation} \label{eq2.08s}
\theta \ge \frac{(4\pi)^3R_m^4k_B T_o\delta_{min}}{P_tG_tG_r\lambda^2\sigma T}. \end{equation}
This formulation ensures radar performance aligns with detection requirements, providing a comprehensive approach to balancing system parameters for enhanced sensing accuracy.

\section{Problem formulation}
\label{S3}
The goal of this proposed optimization problem is to maximize the sum capacity of sensing and communications modes.

\begin{equation}
\label{eq2.09}
\begin{aligned}
      \underset{\theta}{\text{max}} \quad &
     \frac{1}{\ln2}\ln\left(1+\gamma_s \theta \right),
 \\
    \text{subject to} \quad 
                            & \mathcal{C}_1: \frac{\left(1-\theta\right)TB_c}{\ln2}\ln\left(1+\gamma_c\right)\ge R_c \\
                          & \mathcal{C}_2: 0 \le \theta \le 1  \\
                          & \mathcal{C}_3:     \theta\ge\frac{(4\pi)^3 R_m^4k_BT_oB_s\delta_{min}}{P_tG_tG_r\lambda^2\sigma T}
\end{aligned}
\end{equation}
where $\mathcal{C}_1$ ensures the amount of communications data is above a certain cutoff value $R_c$.
$\mathcal{C}_1$ ensures that the amount of communication data transmitted exceeds a specified quality of service (QoS) threshold, $R_c$.
$\mathcal{C}_2$ restricts the variable $\theta$ to lie within the interval $[0, 1]$, while $\mathcal{C}_3$ enforces the minimum radar SNR requirement.
The objective function in (\ref{eq2.09}) is concave, and all constraints are affine, resulting in a convex optimization problem.

To determine the feasible region, we manipulate constraint \( \mathcal{C}_1 \) as follows:
\begin{equation}    
\theta \le \theta_{\max} \triangleq 1 - \frac{R_c \ln 2}{T B_c \ln(1 + \gamma_c)}.
\end{equation}

Therefore, the feasible set for \( \theta \) is:
\begin{equation}
\theta \in \left[\theta_{\min},\; \min(1,\theta_{\max}) \right],
\end{equation}
which is nonempty if \( \theta_{\min} \le \min(1,\theta_{\max}) \).

Since the objective is strictly increasing in \( \theta \), the optimal solution occurs at the upper boundary of the feasible interval. Thus, the closed-form optimal solution is given by:
\begin{equation}
\label{eq:optimal_theta}
\theta^* = 
\begin{cases}
\min(1, \theta_{\max}), & \text{if } \theta_{\min} \le \min(1, \theta_{\max}), \\
\text{Infeasible}, & \text{otherwise}.
\end{cases}
\end{equation}
The optimal time allocation \( \theta^* \) can be used to evaluate the system's power efficiency, defined as
\begin{equation}
    \eta = \frac{\ln\left(1+\gamma_s \theta^* \right) \theta^* + \ln\left(1+\gamma_c\right) (1 - \theta^*)}{
    {\ln2} \left( P_s \theta^* + P_c (1 - \theta^*)\right)}.
\end{equation}
\section{Numerical Results}
\label{S4}
In this section, we present the numerical results of the optimal time allocation for a UAV-enabled ISaC system.
The simulation parameters for sensing and communication are summarized in Table \ref{Table I}.
These are typical values for an airborne ground surveillance radar operating in the X-band \cite{milias2023uas}. 
Whereas, the communications parameters were selected to match the 5G NR numerology \cite{ali2021urllc}.

\begin{table}[!t]
\centering
\caption{Simulation Values.}
\fontsize{10}{12}\selectfont % Adjust font size to 8pt with line skip 10pt
\begin{tabular}{|c||c|c|}
%{|>{\centering\arraybackslash}m{44mm}|m{25mm}|m{25mm}|}
\hline
\textbf{Parameter} &\textbf{Symbol}&\textbf{Value}\\
\hline
\hline
Communications power & $P_c$ & 30 dBm \\
\hline
Sensing power & $P_s$ & 40 dBm \\
\hline
Sensing bandwidth & $B_s$ & 45 MHz\\
\hline
Communications bandwidth & $B_c$& 180 KHz\\
\hline
Pathloss exponent & $\eta$ & 2\\
\hline
Radar cross section (RCS) & $\sigma$ & 20 $\text{m}^2$\\
\hline
Carrier frequency & $f$ & $10.05$ GHz\\
\hline
Maximum velocity & $V_m$ & $30$ m/sec\\
\hline
Antenna gain & $G$ & 8 dB\\
\hline
Time frame length & $T$ & $1$ ms\\
\hline
Ambient temperature & $T_o$ & $300$ K\\
\hline
\end{tabular}
\label{Table I}
\end{table}

\begin{figure}[!t]
\centering

\begin{subfigure}[!t]{0.99\linewidth}%[!t]{0.5\textwidth}
  \centering
  \includegraphics[width=\columnwidth]{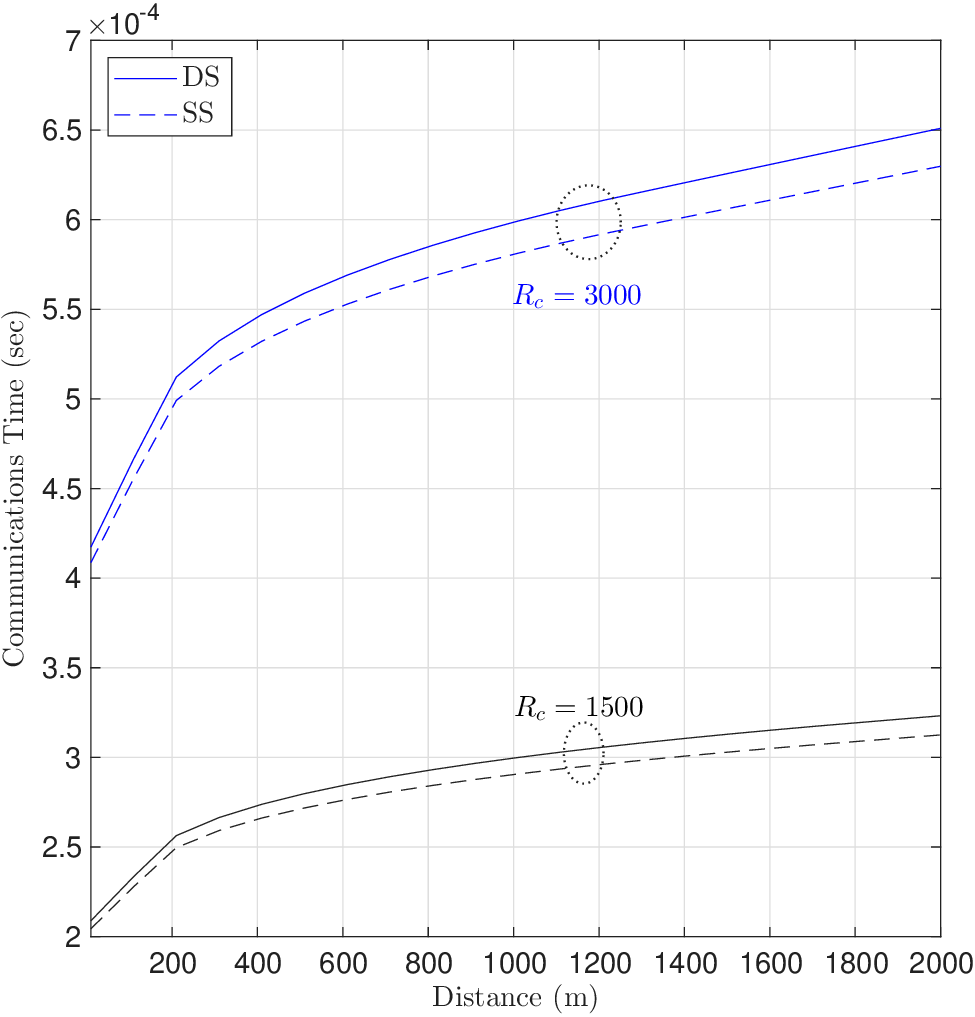}
  \caption{Communications Time}
  \label{fig206_A}
\end{subfigure}\\
\begin{subfigure}[!t]{0.99\linewidth}%[!t]{0.5\textwidth}
  \centering
  \includegraphics[width=\columnwidth]{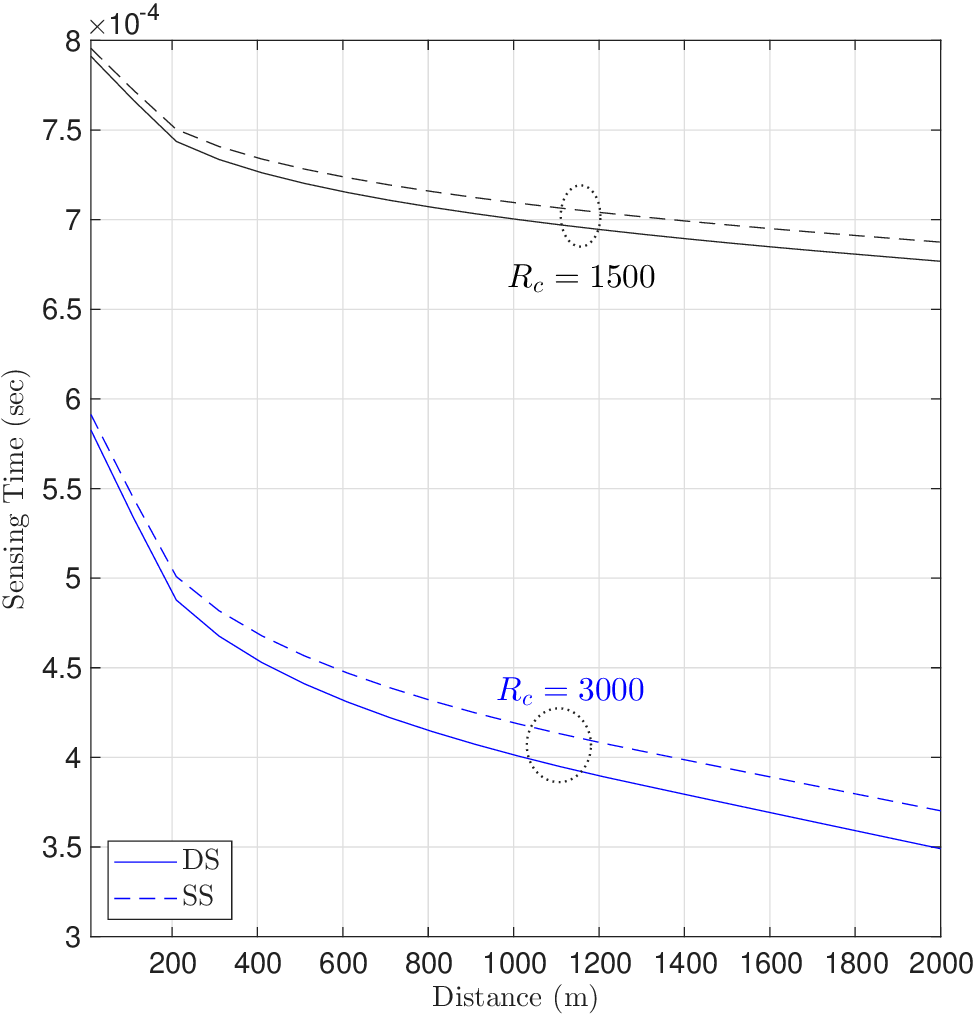}
  \caption{Sensing Time}
  \label{fig206_B}
\end{subfigure}
\caption{Optimal communication and sensing time vs. distance.}
\label{fig206}
\end{figure}

Fig. \ref{fig206} illustrates how the optimal durations for sensing and communication vary with distance $d$, which depends on the UAV’s altitude.
The distance $d$ ranges from $50$ m to $2000$ m, taking into account the elevation of the UAV-based ISaC system.
The channel parameters were selected as ($m_1={\alpha}_1=2.1$, $m_2=m_1+0.3$  and $\alpha_2=\alpha_1+0.3$).
As shown in Fig. \ref{fig206}, the optimal communication time increases with distance to maintain a required data rate. 
Due to the absence of one of the shadowing zones, the SS case requires less communications time, compared to the DS one.
Conversely, the optimal sensing time decreases as the distance $d$ increases.
As expected, increasing the required threshold from 1500 to 2500 bits/frame will require more time allocated for the communications.

\begin{figure}[!t]
\includegraphics[width=\columnwidth]{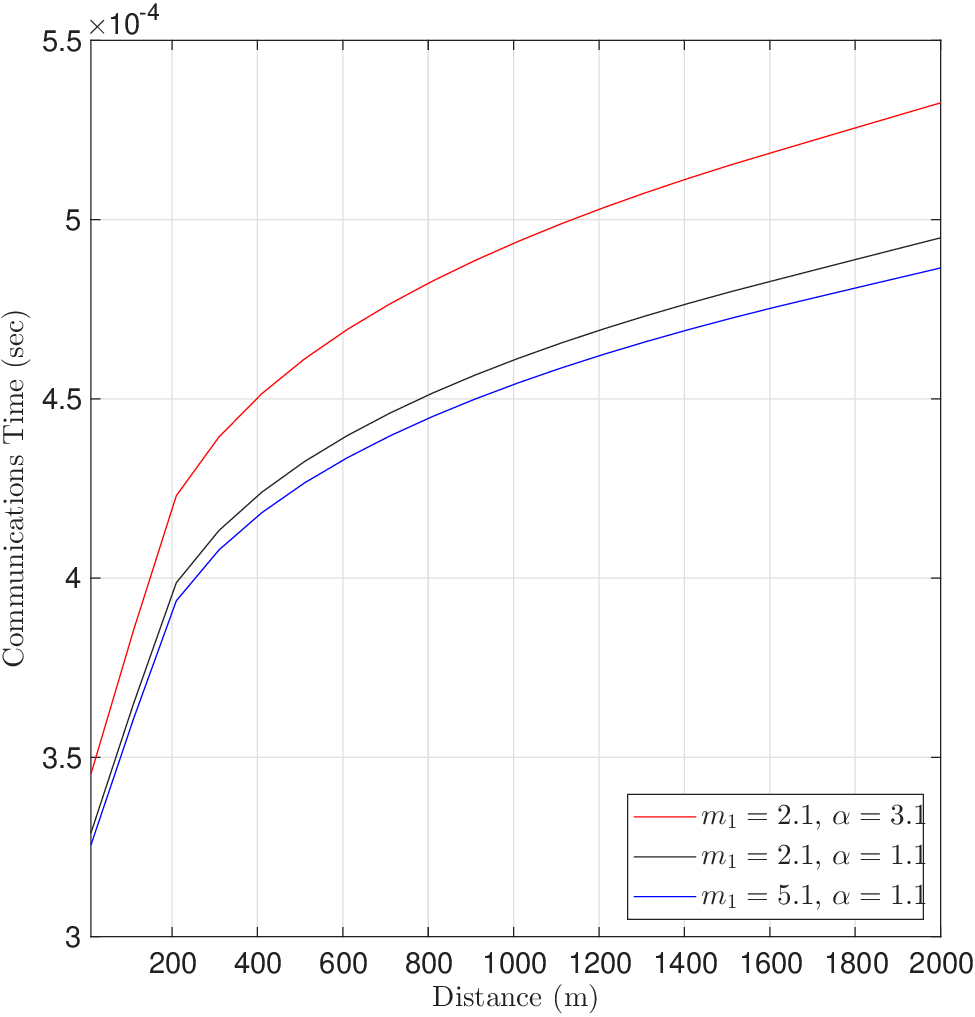}
\caption{Optimal communication time vs. distance (SS-case).}
\label{fig207}
\end{figure}

In Fig. \ref{fig207}, the optimal communication time is shown as a function of distance for the SS case under varying fading parameters, with \( R_c = 2700 \) bits/frame and \( m_2 = m_1 + 0.3 \).
As \( \alpha \) increases from 1.1 to 3.1, the shadowing effect intensifies, degrading the channel gain and necessitating a longer communication time to meet the threshold.
In contrast, increasing \( m_1 \) from 2.1 to 5.1 mitigates multipath effects, enhancing the channel gain and reducing the required communication time.

\begin{figure}[!t]
\centering
\includegraphics[width=\columnwidth]{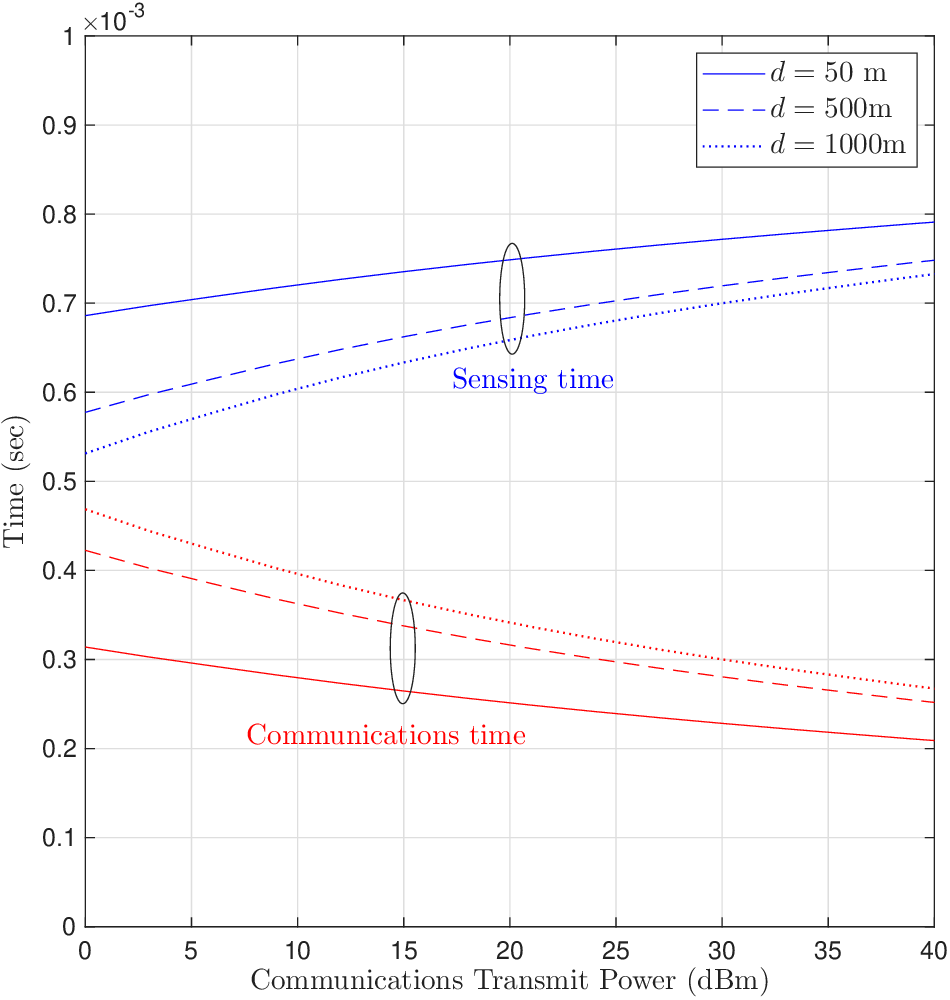}
\caption{Optimal communication and sensing time vs. communications transmit power.}
\label{Figure 210}
\end{figure}

In Fig. \ref{Figure 210},  the optimal communication time is shown as a function of the communications transmit power in dBm for the SS case.
The channel parameters were selected as ($m_1={\alpha}_1=2.1$ and $m_2=2.4$).
At lower transmission power levels, an extended communication duration is required to satisfy the desired QoS requirements, with \( R_c = 2500 \) bits/frame.
This relationship arises from the inverse proportionality between transmission power and communication time for a given data volume, wherein an increase in power leads to a reduction in the required transmission duration.

\begin{figure}[!t]
\centering
\includegraphics[width=\columnwidth]{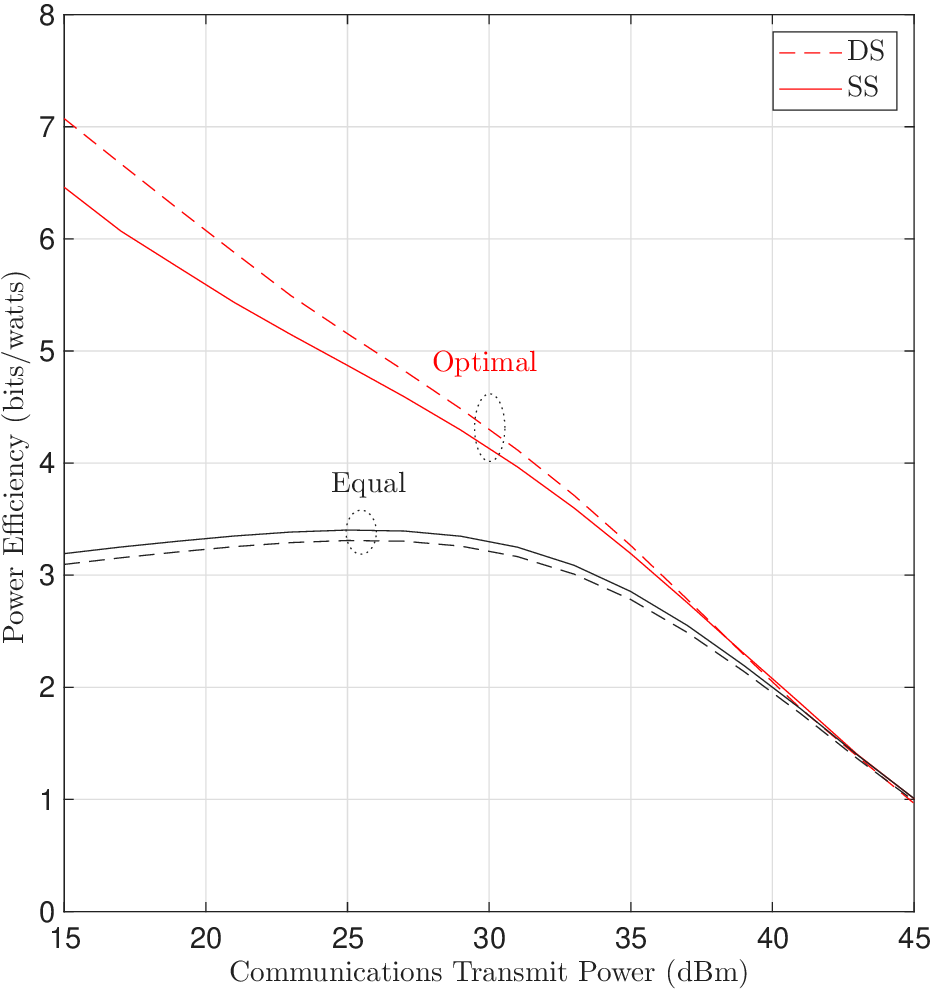}
\caption{Power efficiency vs. communications transmit power.}
\label{Figure 211}
\end{figure}

Fig. \ref{Figure 211} illustrates the power efficiency (bits/watt) as a function of communication transmit power for both DS and SS schemes, under equal and optimal time allocation strategies.
The channel parameters were selected as ($m_1={\alpha}_1=2.1$, $m_2=m_1+0.3$  and $\alpha_2=\alpha_1+0.3$).
As anticipated, increasing the communication transmit power leads to a decrease in power efficiency. The optimal allocation consistently outperforms the equal allocation, particularly in the low and mid-power range (0–35 dBm), highlighting the advantages of jointly optimizing the time split between radar and communication functions.
Moreover, the optimal allocation’s adaptability allows it to achieve higher power efficiency for the DS scheme compared to the SS scheme, at the same communication transmit power level.
At higher transmit powers, the performance of all schemes converges, suggesting diminishing returns from optimization in power-inefficient regions.
At higher powers, the performance of all schemes converges, indicating limited gains from optimization in power-inefficient regions.
This behavior arises because, in the high-power regime, communication throughput saturates due to channel capacity limits, while power consumption continues to increase, leaving little room for improvement through time allocation.

\section{Conclusion}
\label{S5}
This paper optimized time allocation in a UAV-enabled ISaC system operating in the X-band, balancing communication reliability and sensing accuracy under practical fading conditions.
By incorporating both SS and DS models, we captured realistic UAV-to-ground channel variations.
Our results reveal that optimal time partitioning dynamically adapts to distance, fading severity, and data rate requirements, ensuring efficient resource utilization.
These insights contribute to the design of energy-efficient ISaC frameworks, paving the way for enhanced UAV-based wireless networks.
Future research may investigate the impact of UAV mobility and interference from multiple UAVs on ISAC performance.
%\bibliographystyle{IEEEtran}
%\bibliography{bibl.bib}
%\vfill


\end{document}